\documentclass{jfm}

\usepackage{graphicx}
\usepackage{natbib}
\usepackage[usenames,dvipsnames]{xcolor}	

\ifCUPmtlplainloaded \else
  \checkfont{eurm10}
  \iffontfound
    \IfFileExists{upmath.sty}
      {\typeout{^^JFound AMS Euler Roman fonts on the system,
                   using the 'upmath' package.^^J}%
       \usepackage{upmath}}
      {\typeout{^^JFound AMS Euler Roman fonts on the system, but you
                   dont seem to have the}%
       \typeout{'upmath' package installed. JFM.cls can take advantage
                 of these fonts,^^Jif you use 'upmath' package.^^J}%
       \providecommand\upi{\pi}%
      }
  \else
    \providecommand\upi{\pi}%
  \fi
\fi

\ifCUPmtlplainloaded \else
  \checkfont{msam10}
  \iffontfound
    \IfFileExists{amssymb.sty}
      {\typeout{^^JFound AMS Symbol fonts on the system, using the
                'amssymb' package.^^J}%
       \usepackage{amssymb}%
         \let\leq=\leqslant
         \let\geq=\geqslant
      }{}
  \fi
\fi

\ifCUPmtlplainloaded \else
  \IfFileExists{amsbsy.sty}
    {\typeout{^^JFound the 'amsbsy' package on the system, using it.^^J}%
     \usepackage{amsbsy}}
    {\providecommand\boldsymbol[1]{\mbox{\boldmath $##1$}}}
\fi

\newcommand\Rey{\mbox{\textit{Re}}}  

\newsavebox{\astrutbox}
\sbox{\astrutbox}{\rule[-5pt]{0pt}{20pt}}

\newcommand{\e}[1]{\ensuremath{\times 10^{#1}}}
\newcommand{\pd}[2]{\partial_{#1}{#2}} 
\newcommand{\avg}[1]{\ensuremath{\overline{#1}}}         
\newcommand{\Rbl}[2]{\ensuremath{\mathcal{R}^{#1}_{#2}}} 
\newcommand{\Lbl}[2]{\ensuremath{\mathcal{L}^{#1}_\mathrm{BL}(#2)}}
\newcommand{\Lc}[1]{\ensuremath{\mathcal{L}_c(#1)}}
\newcommand{\eqref}[1]{(\ref{#1})}

\title[Marginally stable and turbulent boundary layers in TC flow]
      {Marginally stable and turbulent boundary layers in low-curvature Taylor--Couette flow}

\author[H. J. Brauckmann and B. Eckhardt]%
   {Hannes J. Brauckmann$^1$ and Bruno Eckhardt$^{1,2}$%
\thanks{Email address for correspondence: bruno.eckhardt@physik.uni-marburg.de}}

\affiliation{$^1$Fachbereich Physik, Philipps-Universit\"at Marburg,
Renthof 6, D-35032 Marburg, Germany\\[\affilskip]
$^2$J.M. Burgerscentrum, Delft University of Technology, Mekelweg 2, 2628 CD Delft, The Netherlands}

\pubyear{2010}
\volume{650}
\pagerange{119--126}
\date{\today}

\begin{document}

\maketitle

\begin{abstract}
Marginal stability arguments are used to describe the rotation-number dependence of torque in Taylor--Couette (TC) flow
for radius ratios $\eta \geq 0.9$ and shear Reynolds number $\Rey_S=2\e{4}$. 
With an approximate representation of the mean profile by piecewise linear functions,
characterized by the boundary-layer thicknesses at the inner and outer cylinder and the angular momentum in the center,
profiles and torques are extracted from the requirement that the boundary layers represent marginally stable 
TC subsystems and that the torque at the inner and outer cylinder coincide. This model then 
explains the broad shoulder in the torque as a function of rotation number near $R_\Omega\approx 0.2$. For 
rotation numbers $R_\Omega < 0.07$ the TC stability conditions predict boundary layers in which 
shear Reynolds numbers are very large. Assuming that the TC instability is bypassed by some shear
instability, a second maximum in torque appears, in very good agreement with numerical simulations. The
results show that, despite the shortcomings of marginal stability theory in other cases, it
can explain 
quantitatively the non-monotonic torque variation with rotation number for both the broad maximum as
well as the narrow maximum.
\end{abstract}


\section{Introduction}
In shear flows, hydrodynamic instabilities drive vortical motions that transport momentum between the moving walls,
thereby increasing the drag and the forces needed to move the walls.  
We here investigate this general connection between hydrodynamical instabilities and the resulting driving force (or torque) 
for the case of the flow between two concentric independently rotating cylinders, the Taylor--Couette (TC) flow.
TC flow is also a convenient model in which
to study the effect of rotation on shear turbulence, 
since both shear and rotation can be independently controlled by the differential and the mean rotation of the cylinders, respectively.
The mean rotation is known to influence the stability of TC flow  \citep{Taylor1923,Chandrasekhar1961,Esser1996,Dubrulle2005} 
as well as the torque.  
In TC systems with a ratio of the inner to the outer radius $\eta=r_i/r_o$ of $0.5$ and $0.7$, 
the torque as a function of the mean rotation features a maximum 
that occurs for counter-rotating cylinders \citep{Paoletti2011,VanGils2011,Brauckmann2013,Ostilla2013,Merbold2013}. 
The emergence of this torque maximum was rationalised by the occurrence of intermittent turbulent bursts near the outer cylinder \citep{VanGils2012,Brauckmann2013a}, 
which result from a stabilisation of the outer fluid layer \citep{Chandrasekhar1961}.
Recent numerical simulations revealed that the bursting behaviour disappears 
when the cylinder radii become large in the limit $\eta\rightarrow 1$, 
and simultaneously a new rotation dependence of the torque emerges for $\eta\geq0.9$ \citep{Brauckmann2016a}: 
In this low-curvature TC flow, the torque at a shear Reynolds number of $2\e{4}$ shows two coexisting maxima, 
a broad and a narrow one. 
Moreover, the mean angular momentum profiles were found to have a universal shape 
as long as the outer region was not stabilized by counter-rotation of the cylinders \citep{Brauckmann2016a}.
The aim of the present paper is to predict the rotation dependence of the torque for $\eta\geq0.9$ from a simplified model, 
to explain the origin of the two torque maxima and to rationalise the mean profile shapes 
and their variation with system rotation.

In the development of our model, we are guided by the following considerations.
The mean rotation of the TC system causes a centrifugal instability that drives vortical flows 
which redistribute angular momentum radially.
As a result, the mean profile becomes flat in the centre  
and has higher angular velocity gradients in the boundary layers (BLs) close to the cylinder walls. 
Consequently, the torque, which is proportional to the wall shear stress, rises above its laminar value. 
This description clearly illustrates that instability mechanisms, mean velocity profiles and torques are closely connected.
The connection is most explicit in marginal stability theory, 
initially described for thermal convection \citep{Malkus1954b,Howard1966}, and 
later extended to channel flow \citep{Malkus1956,Malkus1983,Reynolds1967,Goldshtik1970} 
and TC flow with stationary outer cylinder \citep{King1984,Marcus1984b,Barcilon1984}. 
We will here present an extension of previous TC studies to the case of
independently rotating cylinders, and will focus in particular on the rotation dependence of the torque, with
its characteristic non-monotonic behaviour. 
Moreover, we will benchmark the model against results from numerical simulations of TC flow.
As we will see, marginal stability based on TC flows alone is not sufficient to explain all features of the torque curves,
and we will formulate a suitable extension that covers the entire range of rotation numbers.

The paper is organised as follows. 
In~\S \ref{sec:DNS} we define the control parameters, describe the numerical method and present the simulation results. 
These include the rotation dependence of the torque (\S \ref{sec:torque}) 
as well as the shape of angular momentum profiles (\S \ref{sec:profiles}), 
which we both aim to understand by the subsequent modelling in~\S \ref{sec:model}. 
We test to which extent the marginal stability assumptions of the model rationalise the rotation dependence of torque and profiles, 
by comparing model predictions to simulation results in~\S \ref{sec:comparison}. 
Discrepancies between model and numerical results point to a change in the BL dynamics that is analysed in~\S \ref{sec:BLtrans}. 
We conclude with a brief summary and further discussions.

\section{Numerical results}
\label{sec:DNS}
We investigate the motion of an incompressible fluid between two concentric cylinders. 
The flow is driven by rotating the inner and outer cylinder with angular velocities $\omega_i$ and $\omega_o$, respectively. 
We are here interested in the limit of large cylinder radii $r_i$ and $r_o$, 
so that the curvature is small and the radius ratio $\eta=r_i/r_o$ is close to one. 
In this low-curvature limit, the cylinder motion is often described by two parameters 
that can be generalized also to other rotating shear flows  \citep{Nagata1986,Dubrulle2005}: 
the average rotation of the system  and the shear from the differential rotation of the cylinders.
Following \citet{Dubrulle2005}, 
we describe the motion in a reference frame rotating with the mean angular velocity 
$\Omega_\mathrm{rf}=(r_i\omega_i+r_o\omega_o)/(r_i+r_o)$, 
so that the cylinders move with the same 
speed but in opposite directions. 
In this reference frame, the velocity difference between the cylinder walls becomes 
\begin{equation}
  U_0=\frac{2}{1+\eta} \left(r_i\omega_i - \eta r_o\omega_o\right) 
  \label{eq:U0}
\end{equation}
and serves as the characteristic velocity scale. 
The velocity $U_0$ and system rotation $\Omega_\mathrm{rf}$ 
enter the definition of two dimensionless control parameters, the shear
Reynolds number
\begin{equation}
  \Rey_S=\frac{U_0\,d}{\nu}=\frac{2}{1+\eta}\left(\Rey_i-\eta \Rey_o\right)
  \label{eq:Res}
\end{equation}
and the rotation number
\footnote{Note that the sign of $R_\Omega$ is opposite to the definition in \citet{Dubrulle2005}.}
\begin{equation}
  R_\Omega=\frac{2\Omega_\mathrm{rf}\,d}{U_0} = (1-\eta)\frac{\Rey_i+\Rey_o}{\Rey_i-\eta\Rey_o} .
  \label{eq:Reom}
\end{equation}
Here, $d=r_o-r_i$ denotes the gap width between the cylinders and $\nu$ the kinematic viscosity of the fluid. 
Equation~\eqref{eq:Res} also gives the relation of $\Rey_S$ and $R_\Omega$ to the traditional Reynolds numbers 
$\Rey_i=r_i\omega_i d/\nu$ and $\Rey_o=r_o\omega_o d/\nu$ of the inner and outer cylinder.
In the following, all results are rendered dimensionless using advective units, where
the velocity difference $U_0$ from~\eqref{eq:U0} and the gap width $d$ serve as characteristic scales for velocities and lengths, respectively.

To analyse the effect of the system rotation on the turbulence, 
we performed direct numerical simulations (DNS) of TC flow at three shear rates $\Rey_S=5\e{3}$, $10^{4}$ and $2\e{4}$, 
and for various values of the rotation number in the range $-0.1\leq R_\Omega\leq 0.95$. 
However, we focus on $\Rey_S=2\e{4}$, and on radius ratios $\eta \gtrsim 0.9$, 
represented by the two extreme cases $\eta=0.9$ and $\eta=0.99$.
For our simulations we used the spectral code described by \citet{Meseguer2007}, 
which expands the velocity components by Chebyshev polynomials in the radial direction and by Fourier modes in the azimuthal 
and axial direction.
Consequently, the simulated flow is axially periodic, 
and we chose an axial length of $L_z=2$, 
which suffices to represent one pair of counter-rotating Taylor vortices. 
In addition, the azimuthal length of the domain was 
$L_\varphi=3.98$ and $L_\varphi=6.31$ (with periodic boundary conditions) 
for $\eta=0.9$ and $\eta=0.99$, respectively. 
As discussed in \citet{Brauckmann2013}, 
the restriction to only one Taylor vortex pair and the reduced azimuthal length 
have little effect on the torque computation.
In the cases of strongly co-rotating cylinders, we performed the simulations in a reference frame that rotates with $\Omega_\mathrm{rf}$.

The spatial resolution, determined by the highest mode order $(M_r, M_\varphi, M_z)$ in each direction, 
is chosen so that the relative amplitude of the highest mode in each direction drops to $\sim 10^{-4}$. 
This condition was identified as one criterion for a converged torque computation \citep{Brauckmann2013} 
and is achieved in the simulations at $\Rey_S=2\e{4}$ by the resolutions $(M_r, M_\varphi, M_z)=(70,94,94)$ and $(70,158,94)$ for $\eta=0.9$ and $\eta=0.99$, respectively. 
Moreover, the simulations meet two additional convergence criteria identified by \citet{Brauckmann2013}: 
Agreement of the torques at the inner and outer cylinder to within $0.5 \%$ 
and fulfilment of the balance between energy input and dissipation to within $1 \%$.
Most of the DNS data used here are taken from \citet{Brauckmann2016a}, 
with additional computations added in the range $R_\Omega>0.6$.

\subsection{Rotation dependence of the torque}
\label{sec:torque}
\begin{figure}
  \centerline{\includegraphics{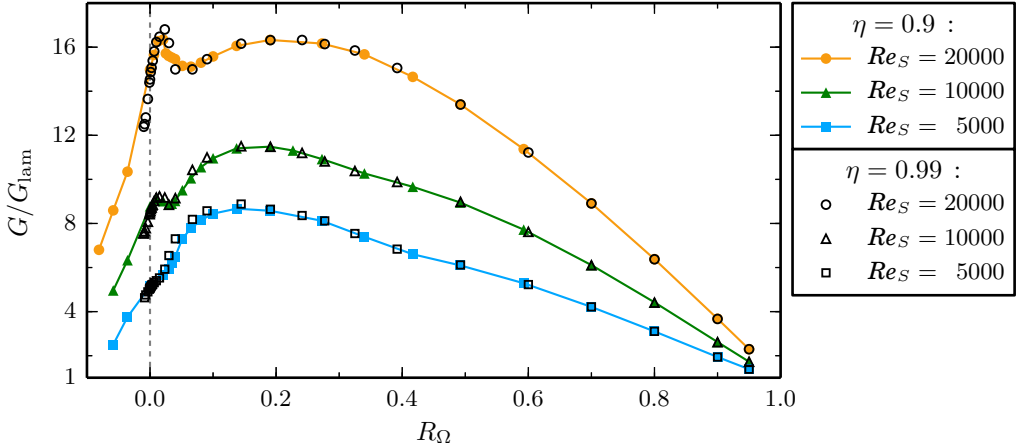}}
  \caption{(Colour online) Variation of the torque $G$ with the system rotation parametrised by $R_\Omega$ for a constant shear $\Rey_S$. 
  In addition to the broad maximum for $\Rey_S=5000$, a second maximum at $R_\Omega=0.02$ emerges with increasing shear. 
  The torque shows the same $R_\Omega$-dependence for both radius ratios $\eta=0.9$ and $\eta=0.99$. 
  All torques are measured in units of the torque $G_\mathrm{lam}$ that would result from a laminar flow.}
  \label{fig:Nu}
\end{figure}

The torque needed to drive the cylinders 
measures the radial transport of angular momentum by the fluid motion \citep{Marcus1984a,Dubrulle2002,Eckhardt2007},
and is strongly influenced by the turbulence in the system.
In the simulations, we calculate the dimensionless value $G=T/(2\upi L_z \rho_f \nu^2)$ of the torque $T$
at the inner and outer cylinder, 
where it is proportional to the mean wall shear stress, 
\begin{equation}
  G= -\Rey_S\, r_x^3 \left.\pd{r}{\avg{\omega}}\right|_{r_x} ,
  \label{eq:G_def}
\end{equation}
with $r_x=r_i$ and $r_x=r_o$ for the inner and outer cylinder, respectively, 
and with the time- and area-averaged angular velocity $\avg{\omega}=\avg{u}_\varphi/r$. 
Here, $\rho_f$ denotes the fluid density.
Figure~\ref{fig:Nu} shows the torque $G$ as a function of $R_\Omega$ at a constant differential rotation $\Rey_S$.
For $\Rey_S=5000$, the torque shows one broad maximum at a rotation number close to $0.2$, 
which in case of $\eta=0.9$ and $\eta=0.99$ corresponds to co-rotating cylinders. 
In contrast, for smaller $\eta$, the torque maximum appears for counter-rotating cylinders 
\citep{Paoletti2011,VanGils2011,Merbold2013,Brauckmann2013} 
and was linked to the occurrence of intermittent turbulent bursts caused by the stabilisation of an outer fluid layer \citep{VanGils2012,Brauckmann2013a}. 
Moreover, low-curvature TC flows with $\eta\geq 0.9$ show
a second, narrower torque maximum at $R_\Omega=0.02$ that 
increases with $\Rey_S$ and becomes similar in magnitude 
to the broad maximum at $\Rey_S=2\e{4}$ \citep{Brauckmann2016a}. 
While this narrow torque maximum occurs for counter-rotating cylinders in the TC system with $\eta=0.9$, 
the rotation number $R_\Omega=0.02$ corresponds to co-rotating cylinders for $\eta=0.99$. 
Consequently, the narrow maximum can not be explained by the intermittent bursts for counter-rotating cylinders \citep{VanGils2012,Brauckmann2013a} 
and, thus, relies on a different mechanism than the torque maximum found for $\eta<0.9$. 
Moreover, the bursting behaviour disappears when $\eta\rightarrow 1$ 
and therefore becomes irrelevant for the torque maximisation for $\eta\gtrsim 0.9$ \citep{Brauckmann2016a}. 
It is worth noting that the $R_\Omega$-dependence of the torque is universal for low-curvature TC flows, 
as demonstrated by the collapse of the torques for $\eta=0.9$ and $\eta=0.99$. 
A similar collapse was also observed in other studies \citep{Dubrulle2005,Paoletti2012,Brauckmann2016a}.
In summary, both torque maxima in low-curvature TC flow 
call for an explanation. 

\subsection{Angular momentum profiles}
\label{sec:profiles}
\begin{figure}
  \centerline{\includegraphics{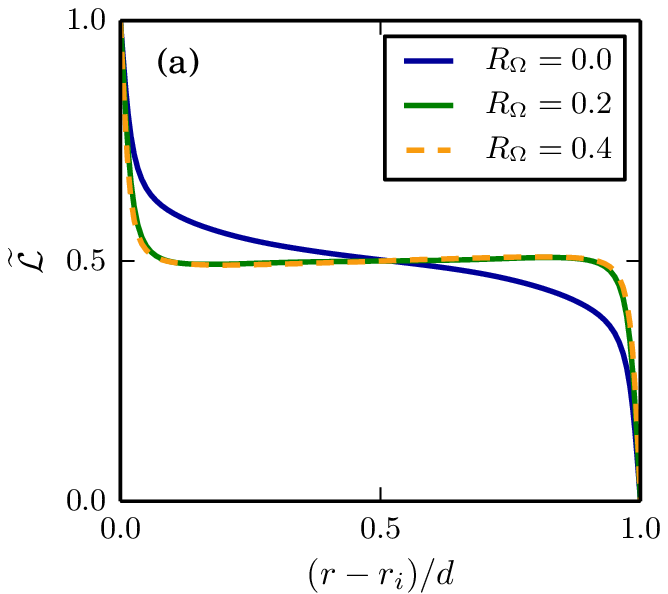}%
              \includegraphics{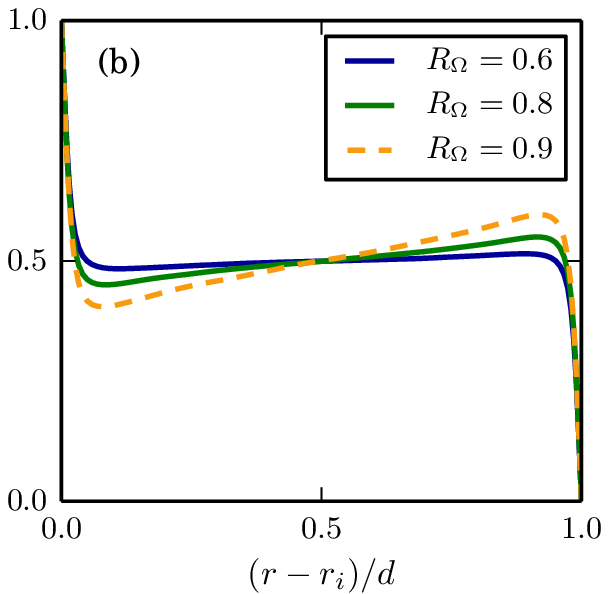}}
  \caption{(Colour online) Angular momentum profiles for various rotation numbers $R_\Omega$ at a constant shear $\Rey_S=2\e{4}$ in a low-curvature TC system with $\eta=0.99$. 
  The time- and area-averaged angular momentum $\avg{\mathcal{L}}$ was rescaled to the interval $(0,1)$ 
  using the transformations $\widetilde{\mathcal{L}}= (\avg{\mathcal{L}}-\mathcal{L}_o) / (\mathcal{L}_i-\mathcal{L}_o)$, 
  where $\mathcal{L}_i$ and $\mathcal{L}_o$ denote the specific angular momentum of the inner and outer cylinder.}
  \label{fig:Lprof}
\end{figure}

An important ingredient to marginal stability considerations are mean profiles
of the specific angular momentum $\mathcal{L}=ru_\varphi$, 
which we obtain by averaging turbulent simulations at $\Rey_S=2\e{4}$ in time and in azimuthal and axial direction.
To exemplify the characteristics of the profiles for low-curvature TC flows, 
figure~\ref{fig:Lprof} shows $\mathcal{L}$~profiles for $\eta=0.99$.
The angular momentum values are rescaled to the interval $(0,1)$ 
to make comparisons between simulations at different $R_\Omega$ easier.
For most rotation numbers, the angular momentum profiles are almost flat in the middle and reach a central value of $\widetilde{\mathcal{L}}\approx0.5$. 
Our recent study \citep{Brauckmann2016a} revealed this profile behaviour also for other radius ratios, 
as long as the flow is not stabilised due to a counter-rotating outer cylinder. 
Flat angular momentum profiles in the centre were also observed in TC experiments 
with the outer cylinder held stationary \citep{Wattendorf1935,Taylor1935,Smith1982,Lewis1999}.

In the limit $\eta\rightarrow 1$, TC flow becomes linearly unstable only in the range $0<R_\Omega<1$ 
for sufficiently high $\Rey_S$ \citep{Dubrulle2005}. 
The $\mathcal{L}$~profile for the lower stability boundary $R_\Omega=0$ 
(corresponding to perfect counter-rotation with $r_i\omega_i=-r_o\omega_o$) 
features a central region of negative slope, see figure~\ref{fig:Lprof}(a). 
Moreover, the gradient of the profile increases 
as $R_\Omega$ tends to $1$, cf.~figure~\ref{fig:Lprof}(b). 
However, this increase is only a consequence of the rescaling of the profiles 
by the difference $(\mathcal{L}_i-\mathcal{L}_o)$: 
In the limit $R_\Omega \rightarrow 1$, this quantity vanishes since the marginal stability boundary $R_\Omega=1$ 
is determined by Rayleigh's criterion 
and the equality of angular momentum at the inner and outer cylinder \citep{Rayleigh1917}.
\begin{figure}
  \centerline{\includegraphics{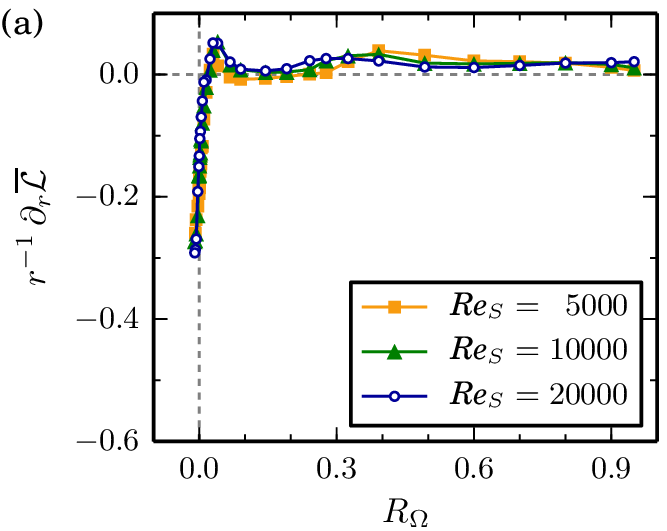}%
              \includegraphics{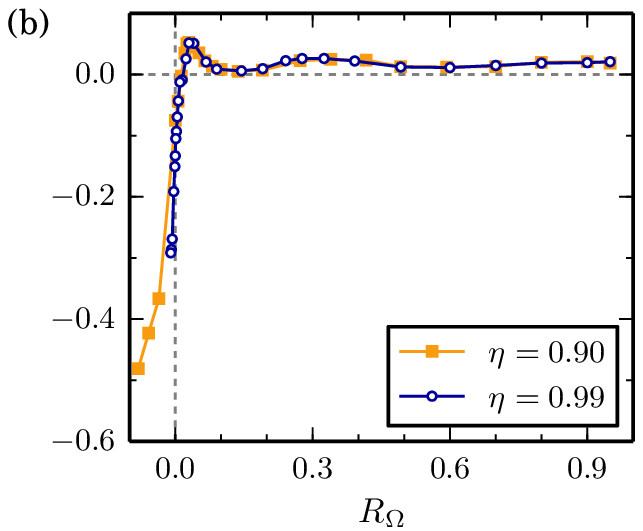}}
  \caption{(Colour online) Angular momentum gradients in the middle region as a function of the rotation number $R_\Omega$. 
  (a)~The gradients for $\eta=0.99$ only slightly vary with $\Rey_S$. 
  (b)~At $\Rey_S=2\e{4}$, the gradients for two different radius ratios coincide. 
  For most values of $R_\Omega$, the angular momentum profiles are almost flat in the middle region with a slight positive slope. 
  We calculated the gradient by averaging the profile derivative divided by the radius ($r^{-1}\pd{r}{\mathcal{\avg{L}}}$) over the central region $(r-r_i)/d\in\left[0.4, 0.6\right]$.}
  \label{fig:Lgrad}
\end{figure}
Indeed, figure~\ref{fig:Lgrad} shows that the profile gradients in the centre measured in advective units do not increase for large $R_\Omega$ 
and are close to zero for $R_\Omega\gtrsim 0$. 
Furthermore, the gradients only slightly vary with $\Rey_S$ (figure~\ref{fig:Lgrad}a) 
and do not differ between simulations for $\eta=0.9$ and $0.99$ (figure~\ref{fig:Lgrad}b), 
which highlights the universal behaviour of the $\mathcal{L}$~profiles in the centre.
It is important to note that a radially constant angular momentum complies with marginal stability according to Rayleigh's inviscid criterion 
and that at high $\Rey_S$, viscosity plays an important role only close to the walls and not in the central region. 
Thus, one can interpret the constant central angular momentum profiles to be in a marginally stable state \citep{Wattendorf1935,Taylor1935,Brauckmann2016a}. 
However, marginal stability is not exactly fulfilled, 
and the $\mathcal{L}$~profiles show a slightly positive slope in the middle 
as previously observed by \citet{Smith1982}, \citet{Lewis1999} and \citet{Dong2007}. 
The general occurrence of an almost flat central region in the angular momentum profile for $R_\Omega\gtrsim 0$ 
will be an important ingredient for the model.

\section{Marginal stability model}
\label{sec:model}
The rotation-number dependence of the torque and of the mean profiles
is a consequence of the complicated turbulent flow that 
is governed by the hydrodynamic equations of motion. 
The shape of the mean profiles, however, can be rationalized by a few simple modelling assumptions, 
as first proposed by \citet{Malkus1954b} and \citet{Howard1966} for the case of thermal convection 
and later applied to TC flow with stationary outer cylinder by \citet{King1984} and \citet{Marcus1984b}. 

\subsection{Defining equations for the model}
The model is based on the following three assumptions:
\begin{figure}
  \centerline{\includegraphics{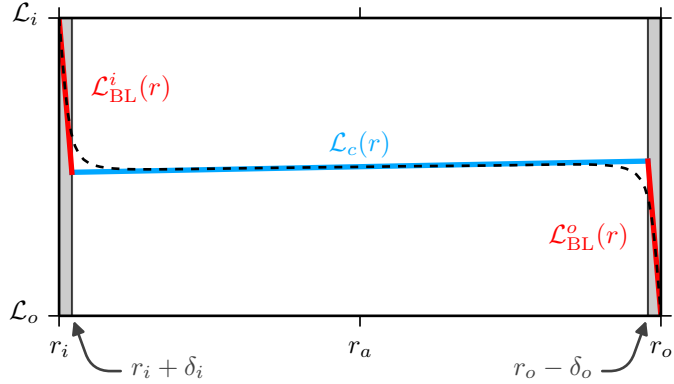}}
  \caption{(Colour online) The marginal stability model describes the angular momentum profile by three linear regions: the inner and outer BL $\Lbl{i}{r}$ and $\Lbl{o}{r}$ having a thickness of $\delta_i$ and $\delta_o$, respectively, and the central region $\Lc{r}$. The model profile (solid lines) calculated for $\eta=0.99$, $\Rey_S=2\e{4}$ and $R_\Omega=0.5$ is compared to the corresponding DNS profile (dashed line).}
  \label{fig:model}
\end{figure}
\renewcommand{\labelenumi}{(\roman{enumi})}%
\begin{enumerate}
	\item 
	The angular momentum profile can be approximated by a sequence of three linear functions as sketched in~figure~\ref{fig:model}: 
	The inner BL profile $\Lbl{i}{r}$ extends from $r_i$ to the radius $r_i+\delta_i$, 
	the outer BL profile $\Lbl{o}{r}$ from the radius $r_o-\delta_o$ to $r_o$ and the central profile $\Lc{r}$ covers the region in between. 
	Here, $\delta_i$ and $\delta_o$ denote the inner and outer BL thickness. 
	The piecewise linear profile approximates the profiles observed in DNS (figure~\ref{fig:Lprof}), 
	but does not capture the smooth transitions between the BLs and the central region. 

	\item
	The numerical simulations show that the 
	angular momentum profile is almost flat in the centre except for 
	a small positive slope, so that it is nearly marginally 
	stable by Rayleigh's criterion for inviscid flows \citep{Rayleigh1917}. 
	We allow for this slope by approximating the profile  
	in the centre with the $\Rey_S$- and $R_\Omega$-independent constant $s=0.02 r_a$. 
	The proportionality to the mean radius $r_a=(r_i+r_o)/2$ accounts for the observation 
	that the profile gradient divided by the radius ($r^{-1}\pd{r}{\avg{\mathcal{L}}}$) 
	is almost independent of $\eta$, $\Rey_S$ and $R_\Omega$ for $R_\Omega\gtrsim0$, cf.~figure~\ref{fig:Lgrad}. 
	Thus, in advective units, the central profile reads
	\begin{equation}
	  \Lc{r} = \mathcal{L}_a + s\left(r-r_a\right) \qquad \mbox{with} \qquad s=0.02 r_a ,
	  \label{eq:Lc}
	\end{equation}
	where $\mathcal{L}_a$ denotes the angular momentum at the mean radius $r_a$. 
	Note that while we fixed the slope $s$, 
	the variable $\mathcal{L}_a$ is as yet unknown and 
	depends on the external parameters $(\eta, \Rey_S, R_\Omega)$. 

	\item
	In analogy to the central region, which is close to being marginally stable by Rayleigh's criterion, we require
	that the BLs are marginally stable, when viewed as TC subsystems that extend only from the 
	walls to the end of the BLs \citep{King1984}. 
These subsystems have one rigid wall (the physical cylinders) and a softer boundary towards the center that 
effectively leaves more space for the instability modes than a rigid wall. 
A similar configuration occurs in TC flow with counter-rotating cylinders 
where the formed vortices are wider than the unstable inner region \citep{Taylor1923}, 
pointing to an increased effective length scale for the instability modes \citep{Donnelly1960,Esser1996}. 
Therefore, we define the effective gap width of the virtual TC systems as $d_i=\widetilde{a}\delta_i$ and $d_o=\widetilde{a}\delta_o$ 
with the constant factor $\widetilde{a}>1$. 
A comparison between model and DNS in~\S \ref{sec:comparison} will reveal 
that $\widetilde{a}=1.5$ represents a reasonable choice in case of $\Rey_S=2\e{4}$.
Therefore, the embedded TC subsystem of the inner BL can be characterised by an effective radius ratio $\eta_i$, 
a first Reynolds number $\Rbl{i}{1}$ for the physical cylinder and a second Reynolds number $\Rbl{i}{2}$ for the BL edge:
\begin{equation}
  \eta_i=\frac{r_i}{r_i+d_i}, \qquad 
  \Rbl{i}{1}=\frac{\widehat{\mathcal{L}}_i d_i}{r_i\nu} , \qquad 
  \Rbl{i}{2}=\frac{\widehat{\mathcal{L}}_c(r_i+d_i)\, d_i}{(r_i+d_i)\nu}, 
  \label{eq:TCi}
\end{equation}
with $d_i=\widetilde{a}\delta_i$. 
Here, the angular momenta $\widehat{\mathcal{L}}_i$ and $\widehat{\mathcal{L}}_c(r_i+d_i)$ are given in physical units and therefore labelled with a hat. 
Since the new unit of length in~\eqref{eq:TCi} is the effective gap width $d_i$, 
the dimensionless radii become $r^i_1=\eta_i/(1-\eta_i)$ and $r^i_2=1/(1-\eta_i)$ for the BL TC system. 
Similarly, the embedded TC subsystem of the outer BL is characterised by an effective radius ratio $\eta_o$, 
a first Reynolds number $\Rbl{o}{1}$ for the BL edge and a second Reynolds number $\Rbl{o}{2}$ for the physical cylinder:
\begin{equation}
  \eta_o=\frac{r_o-d_o}{r_o}, \qquad 
  \Rbl{o}{1}=\frac{\widehat{\mathcal{L}}_c(r_o-d_o)\, d_o}{(r_o-d_o)\nu} , \qquad 
  \Rbl{o}{2}=\frac{\widehat{\mathcal{L}}_o d_o}{r_o\nu}, 
  \label{eq:TCo}
\end{equation}
with $d_o=\widetilde{a}\delta_o$.
Here, the effective gap width $d_o$ is the new unit of length, 
and the dimensionless radii become $r^o_1=\eta_o/(1-\eta_o)$ and $r^o_2=1/(1-\eta_o)$.
\end{enumerate}

The constants $s$ and $\widetilde{a}$ are fixed by empirical observations, 
so that the model has three variables: 
the angular momentum in the centre $\mathcal{L}_a$ and the BL thicknesses $\delta_i$ and $\delta_o$.
They can be fixed and related to the external parameters $(\eta,\Rey_S,R_\Omega)$ by the following considerations:
First, we implement the assumption of marginal stability from~(iii) 
by requiring that both BLs described by the parameters~\eqref{eq:TCi} and~\eqref{eq:TCo} fulfil 
the stability criterion for laminar TC flow, as described by \citet{Esser1996}. 
We resort to their study since they provide analytic expressions for the stability boundary in the full parameter space
which are in good agreement with experimental results.
These two conditions for the BLs can be solved for $\delta_i$ and $\delta_o$, 
with an implicit dependence on $\mathcal{L}_a$, 
which enters the Reynolds numbers $\Rbl{i}{2}$ and $\Rbl{o}{1}$ via the central profile 
$\Lc{r}$ from~\eqref{eq:Lc} evaluated at $r=r_i+d_i$ and $r=r_o-d_o$, respectively.

The third condition needed to fix the parameters follows from the requirement that 
in the statistically stationary state and averaged over long times, 
the torque exerted on the inner cylinder equals that exerted on the outer cylinder.
Since the torque is proportional to the mean wall shear stress, cf. equation~\eqref{eq:G_def}, 
which is calculated from the linearly approximated BL profiles $\Lbl{i}{r}$ and $\Lbl{o}{r}$, 
the dimensionless torques at the inner and outer cylinder read
\begin{subeqnarray}
  G_i&=& \Rey_S\left(-r_i \left.\pd{r}{\mathcal{L}^i_\mathrm{BL}} \right|_{r_i} +2\mathcal{L}_i\right) 
   = \Rey_S \left(r_i\frac{\mathcal{L}_i-\Lc{r_i+\delta_i}}{\delta_i} +2\mathcal{L}_i\right) , \\
  G_o&=& \Rey_S\left(-r_o \left.\pd{r}{\mathcal{L}^o_\mathrm{BL}} \right|_{r_o} +2\mathcal{L}_o\right) 
   = \Rey_S \left(r_o\frac{\Lc{r_o-\delta_o}-\mathcal{L}_o}{\delta_o} +2\mathcal{L}_o\right) .
   \label{eq:Gmodel}
\end{subeqnarray}
Thus, the third condition becomes $G_i=G_o$.
Since all three conditions are coupled, 
and the stability equations given by \citet{Esser1996} are implicit, we solve the equations numerically.
For given parameter values $(\eta,\Rey_S,R_\Omega)$, this procedure results in predictions for $\mathcal{L}_a$, $\delta_i$ and $\delta_o$, 
and thus for the torque via equation~\eqref{eq:Gmodel}.
In this context, it is important to note that the BL thicknesses~$\delta_i$ and~$\delta_o$ are used to approximate 
the profile derivatives in~\eqref{eq:Gmodel}, 
whereas the increased gap widths~$d_i$ and~$d_o$ are relevant for the Reynolds numbers 
in~\eqref{eq:TCi} and~\eqref{eq:TCo} that describe the stability of the BLs.

\subsection{Predictions}
\label{sec:comparison}
 
The predictions from the model can be compared with the quantities calculated from our numerical simulations. 
While the definitions of the central angular momentum $\mathcal{L}_a=\mathcal{L}(r_a)$ 
and of the torque $G$ also apply to the DNS, 
a BL-thickness definition inspired by the model is needed for general $\mathcal{L}$~profiles from simulations or experiments.
Therefore, we also approximate the angular momentum profiles from the DNS by piecewise linear functions 
similar to the model profile in~figure~\ref{fig:model}. 
We define the distance of the two intersection points to the corresponding wall as BL thicknesses $\delta_i$ and $\delta_o$. 
These lines are obtained by a linear fit to the middle region of the DNS profile 
and by using the derivatives $\left.\pd{r}{\avg{\mathcal{L}}}\right|_{r_i}$ and $\left.\pd{r}{\avg{\mathcal{L}}}\right|_{r_o}$ as the 
slope for the segments in the inner and outer BL, respectively.

We study the marginal stability model for a constant shear $\Rey_S=2\e{4}$ 
and for the two radius ratios $\eta=0.99$ and $\eta=0.9$. 
The first $\eta$-value was chosen to analyse the limit where the cylinder curvature plays a negligible role. 
For $\eta=0.99$, the dimensionless cylinder radii become $r_i/d=99$ and $r_o/d=100$, and the inner and outer BL behave similarly. 
Therefore, we only show results for the inner BL here.
On the other hand, $\eta=0.9$ corresponds to the smallest radius ratio 
for which we still observe the two torque maxima that are characteristic of the low-curvature TC flow \citep{Brauckmann2016a}. 
Since the cylinder radii are approximately ten times smaller compared to the $\eta=0.99$ case, 
we expect the curvature to become relevant. 
Thus, the $\eta=0.9$ case enables us to analyse how curvature effects are represented in the marginal stability model.

\begin{figure}
  \centerline{\includegraphics{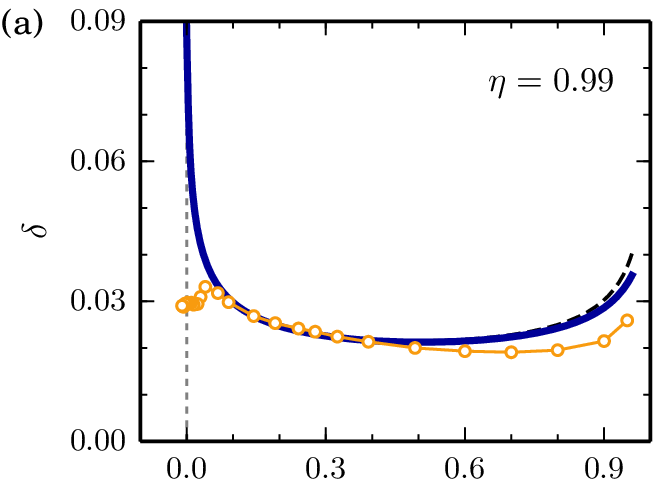}%
	          \includegraphics{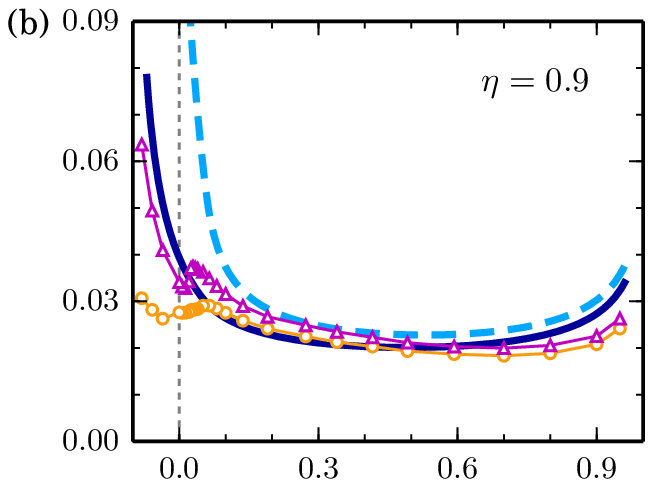}}
  \centerline{\includegraphics{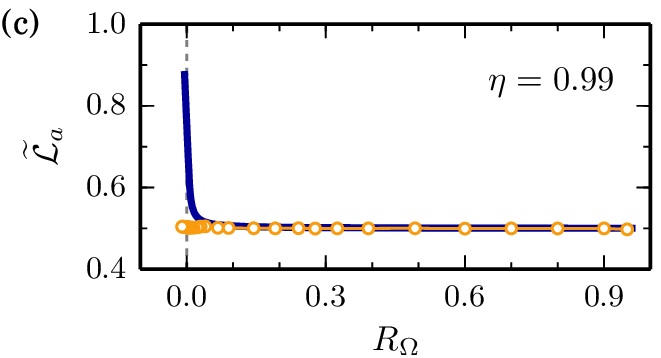}%
		      \includegraphics{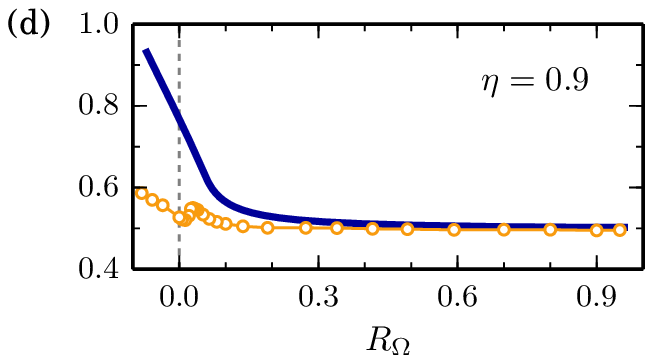}}
  \caption{(Colour online) Comparison of the model predictions (lines) with the DNS results (symbols) for $\Rey_S=2\e{4}$ and two radius ratios 
  showing BL thicknesses in~(a,b) and the central angular momentum $\mathcal{L}_a=\avg{\mathcal{L}}(r_a)$ rescaled by the transformation
  $\widetilde{\mathcal{L}}_a=(\mathcal{L}_a-\mathcal{L}_o)/ (\mathcal{L}_i-\mathcal{L}_o)$ 
  in~(c,d) as a function of $R_\Omega$. 
  Since $\delta_o\approx\delta_i$ for $\eta=0.99$, (a)~only includes the inner BL thickness $\delta_i$ (solid line, circles). 
  In~(b) the outer BL thickness $\delta_o$ is shown by the dashed line (model) and by triangles (DNS). 
  Assuming a constant profile in the centre, i.e. inserting $s=0$ into~\eqref{eq:Lc}, 
  increases the BL thickness predicted by the model for large $R_\Omega$ as illustrated for $\delta_i$ by the black dashed line in~(a).}
  \label{fig:comp-prof}
\end{figure}
Figure~\ref{fig:comp-prof} compares the model prediction for BL thickness and central angular momentum to corresponding DNS results 
and shows the variation of these quantities with the mean system rotation parametrised by $R_\Omega$. 
For $\eta=0.99$, the model and DNS results for~$\delta_i$ coincide in a wide $R_\Omega$ range.
They deviate for $R_\Omega \gtrsim 0.5$, 
but both still show the same upward trend for large $R_\Omega$ (figure~\ref{fig:comp-prof}a). 
When the system rotation $R_\Omega$ tends to zero, the model drastically overestimates the BL thickness.
This discrepancy will be explained and resolved in~\S \ref{sec:BLtrans}.
We observe a similar agreement between model and DNS for $\eta=0.9$ in~figure~\ref{fig:comp-prof}(b), 
where the variation of the inner and outer BL thickness with $R_\Omega$ resembles the $\eta=0.99$ case. 
However, for $\eta=0.9$ the outer BL is thicker than the inner one, 
and the marginal stability model correctly reproduces this curvature effect. 
Furthermore, the curvature causes a radial difference in stability when the cylinders counter-rotate, 
which happens in case of $\eta=0.9$ for $R_\Omega<0.1$. 
Then, the counter-rotating outer cylinder stabilises an outer layer while the inner region is still 
centrifugally unstable \citep{Chandrasekhar1961}. 
Such a stabilisation permits a thicker BL, 
and both DNS and model reflect the radial difference in stability in a~$\delta_o$ 
that is much larger than~$\delta_i$ for negative and slightly positive $R_\Omega$ values.

The model prediction for the central angular momentum $\mathcal{L}_a$ agrees well with the DNS result, 
except for small $R_\Omega$ values, as shown in~figure~\ref{fig:comp-prof}(c,d). 
For most rotation numbers, $\mathcal{L}_a$ reaches the mean value $(\mathcal{L}_i+\mathcal{L}_o)/2$ 
corresponding to $\widetilde{\mathcal{L}}_a=0.5$, 
which indicates that both BLs feature the same angular momentum drop. 
This symmetric behaviour changes in the marginal stability model as shown by the 
increase of $\widetilde{\mathcal{L}}_a$ for $R_\Omega\rightarrow 0$, 
which implies a larger angular momentum difference over the outer BL. 
Together with this profile asymmetry, the aforementioned stabilisation of the outer BL 
caused by counter-rotating cylinders 
enables a larger shear gradient in the outer BL while maintaining marginal stability. 
For $\eta=0.9$, the predicted $\widetilde{\mathcal{L}}_a$ starts to increase at a larger $R_\Omega$ value than for $\eta=0.99$. 
This is in line with the fact that 
counter-rotation corresponds to $R_\Omega<0.1$ and 
$R_\Omega<0.01$ for $\eta=0.9$ and $\eta=0.99$, respectively.

\begin{figure}
	\centerline{\includegraphics{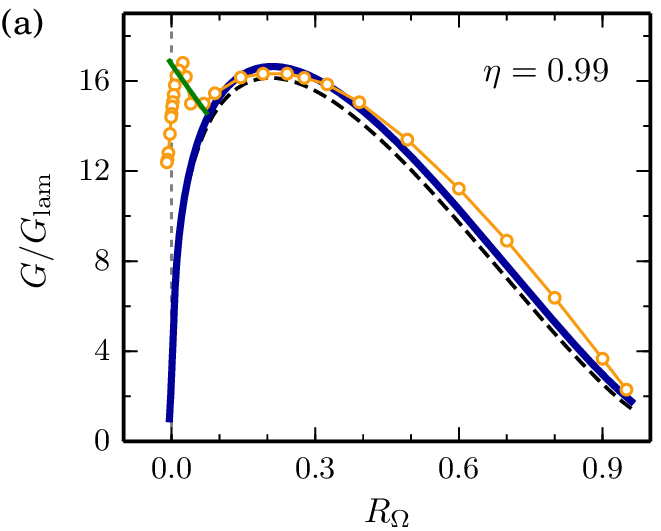}%
	            \includegraphics{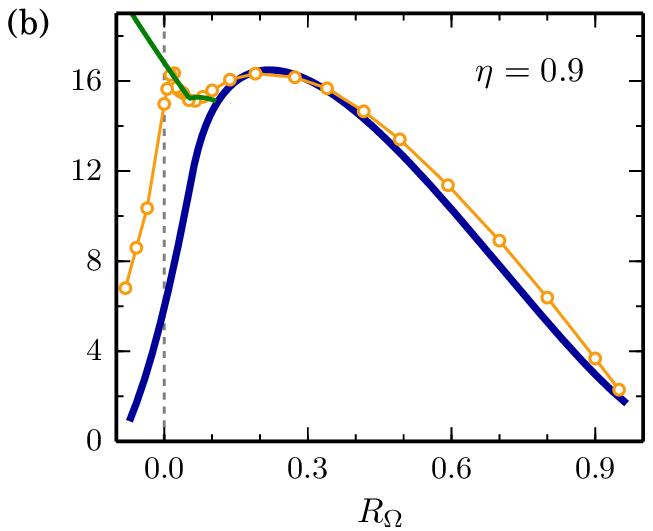}}
  \caption{(Colour online) Comparison of the torques $G$ estimated by the model (dark blue line) and calculated in the DNS (circles) 
  as a function of $R_\Omega$ for $\Rey_S=2\e{4}$ and two radius ratios. 
  The torque is normalised by its laminar value $G_\mathrm{lam}$.  
  The green (grey) line for $R_\Omega<0.07$ in~(a) and for $R_\Omega<0.1$ in~(b) originates from an improved model discussed in~\S \ref{sec:BLtrans} 
  which rationalises the beginning of the narrow maximum. 
  Assuming a constant profile in the centre, i.e. inserting $s=0$ into~\eqref{eq:Lc}, 
  only slightly changes the model prediction for the torque as exemplified by the dashed line in~(a).}
  \label{fig:comp-nu}
\end{figure}

Equation~\eqref{eq:Gmodel} translates these profile characteristics, i.e. $\mathcal{L}_a$, $\delta_i$ and $\delta_o$, 
into a marginal stability prediction for the torque, 
which is compared to the DNS results for $\eta=0.99$ and $\eta=0.9$ in~figure~\ref{fig:comp-nu}. 
Similar to the behaviour of the BL thicknesses (cf.~figure~\ref{fig:comp-prof}a,b), 
the torques coincide in the range $0.1\lesssim R_\Omega\lesssim 0.5$ and show small deviations for larger rotation numbers. 
Thereby, the model reproduces the broad torque maximum at $R_\Omega=0.2$, 
suggesting that the marginal stability of mean profiles  
is responsible for this rotation-number dependence of the torque.
This is apparently different from the case of the magnetorotational instability in TC flows, where the maximum could be related to parameters corresponding to maximal growth rates \citep{Guseva2015}.
The model does not reproduce the narrow torque maximum at $R_\Omega=0.02$ from the DNS, 
but it predicts a strong decrease in $G$ as $R_\Omega$ tends to zero. 
Consequently, the formation of the second torque maximum must result from another mechanism
that will be discussed in the following section.

Finally, we note that the only model parameter 
whose value was not determined by the model assumptions is the constant $\widetilde{a}>1$ 
which describes the effectively larger gap widths $d_i=\widetilde{a}\delta_i$ and $d_o=\widetilde{a}\delta_o$ for the Reynolds numbers of the BLs. 
The introduction of $\widetilde{a}$ is physically justified by the free-surface boundary condition at the BL edge, 
and its value determines the general magnitude of model torques in~figure~\ref{fig:comp-nu}. 
However, the variation of the torque with $R_\Omega$ does not depend 
critically 
on the value of $\widetilde{a}$. 
We chose the constant $\widetilde{a}=1.5$ so that the amplitude of the model-torque maximum matches the DNS torques.
In contrast, the magnitude of the profile slope in the centre was set to $s=0.02 r_a$ beforehand 
in accordance with empirical observations. 
Alternatively, one could have postulated that the central profile exactly realises marginal stability according to Rayleigh's criterion, 
which requires a constant angular momentum and thus the slope $s=0$. 
In a previous marginal stability model, the angular momentum was assumed to be constant in the central region \citep{King1984,Marcus1984b}, 
and the effect of setting $s=0$ in our model is exemplified for $\eta=0.99$ by the dashed line in~figures~\ref{fig:comp-prof}(a) and~\ref{fig:comp-nu}(a): 
The BL-thickness predictions with $s=0$ and $s=0.02 r_a$ only differ for large $R_\Omega$ values, 
with the $s=0.02 r_a$ case being closer to the DNS result. 
Similarly, the model prediction for the torque only slightly varies with the value of $s$, 
and the variation would not be recognisable for the central angular momentum $\mathcal{L}_a$ in~figure~\ref{fig:comp-prof}(c). 
The discussion shows that the value of $\widetilde{a}$ and the 
choices for the profile gradient in the central region have only minor effects on the torque, thereby
demonstrating the robustness of the model.

\section{Boundary-layer transition}
\label{sec:BLtrans}
\begin{figure}
	\centerline{\includegraphics{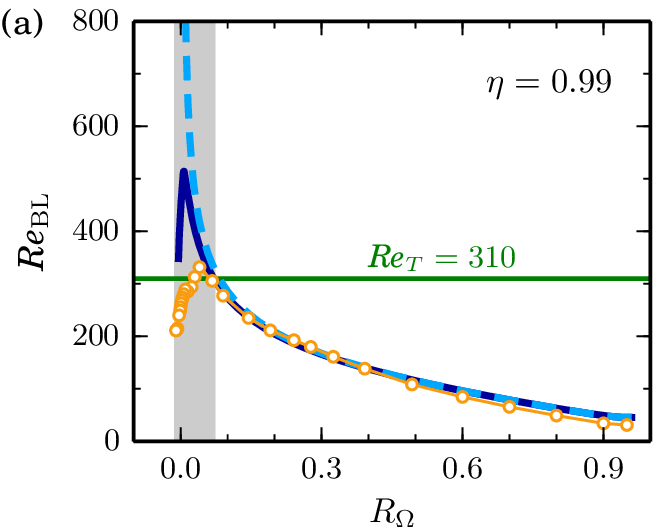}%
	            \includegraphics{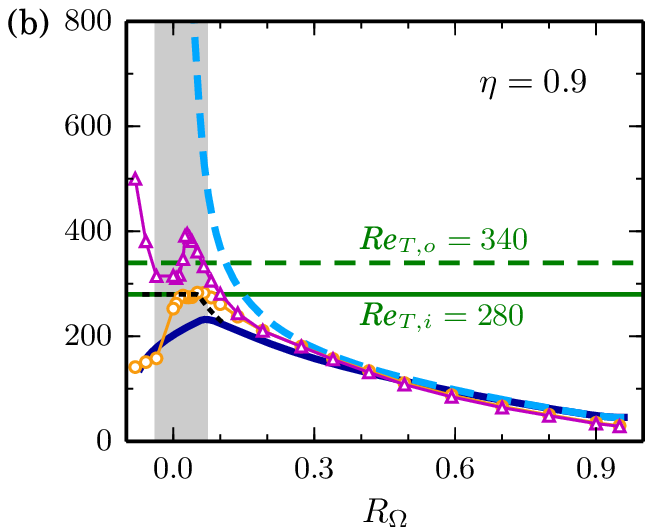}}
  \caption{(Colour online) Inner and outer BL Reynolds number $\Rey_\mathrm{BL}$ 
  predicted by the model (solid and dashed line) and calculated from the DNS (circles and triangles) 
  for $\Rey_S=2\e{4}$ and two radius ratios. 
  In the DNS for $\eta=0.99$, $\Rey^o_\mathrm{BL}$ coincides with $\Rey^i_\mathrm{BL}$ and therefore is left out in~(a). 
  While $\Rey_\mathrm{BL}$ predicted by marginal stability strongly increases for $R_\Omega\rightarrow0$ (only for outer BL in~(b)), 
  it reaches a maximum and drops again in the DNS, indicating a transition to turbulent BLs, 
  which occurs for $R_\Omega<0.07$ in the grey-shaded area. 
  This is implemented in an improved model by additionally requiring 
  that the BL Reynolds number cannot exceed a transition Reynolds number $\Rey_T$ marked by the green (grey) lines. 
  Then, the model reproduces the beginning of the narrow torque maximum in~figure~\ref{fig:comp-nu}.
  In the improved model, $\Rey^i_\mathrm{BL}$ for $\eta=0.9$ also increases to $\Rey_{T,i}$ as shown by the dotted line in~(b).}
  \label{fig:ReBL}
\end{figure}
The observed discrepancy between model and DNS for small rotation numbers points to a deviation 
of the flow from the marginal stability behaviour. 
Since this discrepancy also occurs for the BL thicknesses in~figure~\ref{fig:comp-prof}(a,b), 
we expect that the change in stability takes place in the BLs. 
To further assess their stability, we assign a shear Reynolds number to the inner and outer BL defined as
\begin{equation}
  \Rey^i_\mathrm{BL}= \frac{\hat{r}_i\left(\omega_i-\left.\omega\right|_{r_i+\delta_i}\right)\delta_i}{\nu} , \qquad
  \Rey^o_\mathrm{BL}= \frac{\hat{r}_o\left(\left.\omega\right|_{r_o-\delta_o}-\omega_o\right)\delta_o}{\nu} ,
\label{eq:ReBL}
\end{equation}
with the typical radii $\hat{r}_i=r_i+\delta_i/2$ and $\hat{r}_o=r_o-\delta_o/2$. 
These Reynolds numbers are based on the angular velocity gradient across the BL 
and resemble the ones defined by \citet{VanGils2012}.
Figure~\ref{fig:ReBL} compares $\Rey^i_\mathrm{BL}$ and $\Rey^o_\mathrm{BL}$ predicted by the model (lines) to the corresponding DNS results (symbols). 
For the DNS with $\eta=0.99$, we only show results for $\Rey^i_\mathrm{BL}$ 
since they coincide with $\Rey^o_\mathrm{BL}$. 
While for most rotation numbers, model and DNS are in good agreement, 
the pronounced discrepancy for small $R_\Omega$ values occurs again. 
In the DNS, the shear gradient across the BL increases with decreasing rotation number, 
reaches a maximum at a small positive $R_\Omega$ value and then drops again. 
In contrast, the model predicts a drastic increase of the BL Reynolds number (only $\Rey^o_\mathrm{BL}$ for $\eta=0.9$) when $R_\Omega$ tends to $0$. 
This increase is unrealistic since BLs are known to undergo a transition to turbulence 
if their Reynolds number exceeds a critical value $\Rey_T$ \citep{Schlichting}, 
as previously discussed for TC flow by \citet{VanGils2012}. 
For example, a Prandtl--Blasius BL becomes linearly unstable for $\Rey_\mathrm{BL}>520$ \citep{Schmid}.
However, the presence of free-stream turbulence above the BL 
(as is the case here in TC flow) lowers the transition Reynolds number $\Rey_T$ \citep{vanDriest1963,Andersson1999} 
since such strong disturbances can cause bypass transitions in the BL.
Consequently, the marginal stability of the BLs as determined from the  
TC stability criterion \citep{Esser1996} is 
bypassed by a transition to turbulence following another route \citep{Faisst2000}.

This BL transition can be incorporated into the model by means of the additional assumption 
that the BLs are also marginally stable with respect to a transition Reynolds number $\Rey_T$, 
which means that $\Rey^i_\mathrm{BL}$ and $\Rey^o_\mathrm{BL}$ must equal $\Rey_T$ 
if they would exceed this value otherwise. 
The critical values $\Rey_T=310$ for $\eta=0.99$ as well as $\Rey_{T,i}=280$ and $\Rey_{T,o}=340$ for $\eta=0.9$ approximate the maximal magnitude of the shear gradient occurring in the DNS, 
as indicated by the horizontal lines in~figure~\ref{fig:ReBL}. 
These $\Rey_T$ values suggest that, 
as a result of the increased cylinder curvature for $\eta=0.9$, 
the outer BL becomes turbulent at a higher shear rate than the inner one. 
Furthermore, we note for the $\eta=0.9$ case that in the improved model, 
$\Rey^i_\mathrm{BL}$ also reaches the transition at $\Rey_{T,i}$ as demonstrated by the dotted line in~figure~\ref{fig:ReBL}(b).

With the additional assumption of marginal stability with respect to the BL transition, 
the model now also reproduces the onset of the narrow torque maximum, 
as shown by the green (grey) line for $R_\Omega<0.07$ and for $R_\Omega<0.1$ in~figure~\ref{fig:comp-nu}(a) and~(b), respectively. 
The two bends in the torque curve for $\eta=0.9$ result from two different $R_\Omega$ values for the inner and outer BL transition in this case.
In contrast to the DNS results, the model still does not include the torque decrease for $R_\Omega<0.02$, 
which, however, is plausible since in the limit $\eta \rightarrow 1$, the complete flow becomes linearly stable for $R_\Omega<0$ \citep{Dubrulle2005}. 
In summary, the model suggests that the narrow torque maximum originates from the transition to turbulent BLs 
for rotation numbers $R_\Omega<0.07$ highlighted by shaded regions in~figure~\ref{fig:ReBL}. 

\begin{figure}
	\centerline{\includegraphics{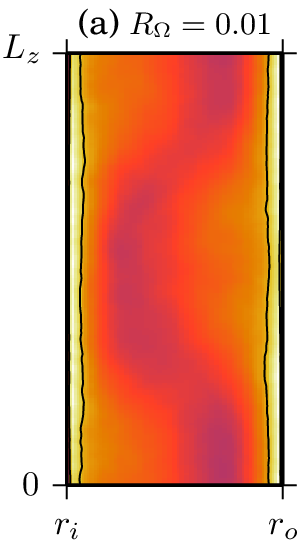}%
	            \includegraphics{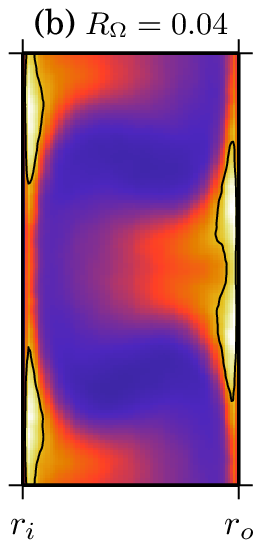}%
	            \includegraphics{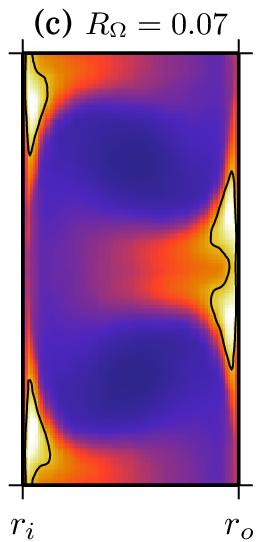}%
	            \includegraphics{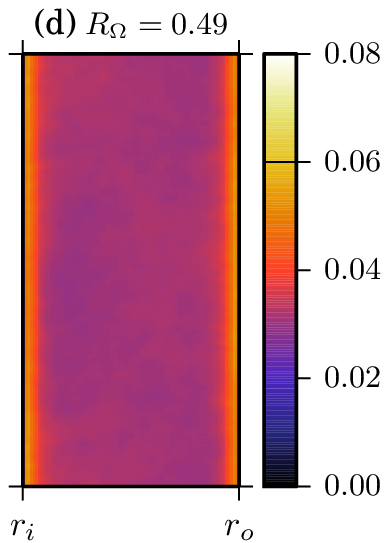}}
  \caption{(Colour online) Azimuthal- and time-averaged RMS ($\mathcal{L}_\mathrm{rms}$) of the angular momentum fluctuations 
  $\mathcal{L}'=\mathcal{L} -\left<\mathcal{L}\right>_{\varphi,t}$ 
  calculated for various values of $R_\Omega$ in DNS with $\eta=0.99$ and $\Rey_S=2\e{4}$. 
  $\mathcal{L}$ is measured in units of $\widetilde{r}U_o$ with the geometric mean radius $\widetilde{r}=\sqrt{r_i r_o}$. 
  All plots use the same colour scale, and the contour line marks strongly turbulent regions with $\mathcal{L}_\mathrm{rms}>0.06$, 
  which are absent in~(d), cover a limited axial fraction of the BL in~(b,c) and extent over the entire BL in~(a). 
  The axial fraction $F_T$ of the strongly turbulent regions is further analysed in~figure~\ref{fig:FT+scal}(a).}
  \label{fig:Lrms}
\end{figure}
Turbulent BLs consist of small vortices that generate high- and low-speed streaks close to the wall, 
which cause strong fluctuations of the downstream velocity $u_\varphi$ and likewise of $\mathcal{L}=ru_\varphi$. 
Remarkably, the angular momentum fluctuations $\mathcal{L}'=\mathcal{L} -\left<\mathcal{L}\right>_{\varphi,t}$ are generally of comparable amplitude in both BL regions 
in contrast to the velocity fluctuations $u_\varphi'$.
Therefore, we analyse the azimuthal- and time-averaged root-mean-square (RMS) 
of the angular momentum fluctuations $\mathcal{L}_\mathrm{rms}= (\left<\mathcal{L}'^2\right>_{\varphi,t})^{1/2}$ 
(see also \citet{Ostilla-Monico2014}). 
Figure~\ref{fig:Lrms} shows $\mathcal{L}_\mathrm{rms}$ in the radial-axial plane 
for various values of $R_\Omega$ and the example case $\eta=0.99$. 
Since all plots use the same colour scale, 
it becomes apparent that for $R_\Omega=0.49$ (figure~\ref{fig:Lrms}d) the fluctuations are relatively small indicating laminar BLs. 
At the same time, no axial variation in $\mathcal{L}_\mathrm{rms}$ that would indicate the presence of Taylor vortices is discernible.
This changes with decreasing $R_\Omega$ as exemplified for $R_\Omega=0.07$ in~figure~\ref{fig:Lrms}(c): 
A limited axial fraction of each BL becomes turbulent, 
as evidenced by strong fluctuations in the regions marked by the contour line at $\mathcal{L}_\mathrm{rms}=0.06$. 
The axial position of these turbulent BL regions correlates with the radial flow produced by the existing Taylor vortex pair: 
Inner and outer BL are turbulent only adjacent to the outflow (top/bottom) and inflow region (middle), respectively. 
The coexistence of laminar and turbulent regions in the BLs corresponds to the \textit{transitional regime} 
described by \citet{Ostilla-Monico2014} for $\eta=0.714$ and a stationary outer cylinder.
When $R_\Omega$ further decreases below $0.07$, 
the turbulent part of each BL grows in height (figure~\ref{fig:Lrms}b) until the entire BL becomes turbulent (figure~\ref{fig:Lrms}a), 
as suggested by the marginal stability model. 
Simultaneously, the axial variation of $\mathcal{L}_\mathrm{rms}$ in the centre and, hence, the Taylor vortices become weaker.

\begin{figure}
	\centerline{\includegraphics{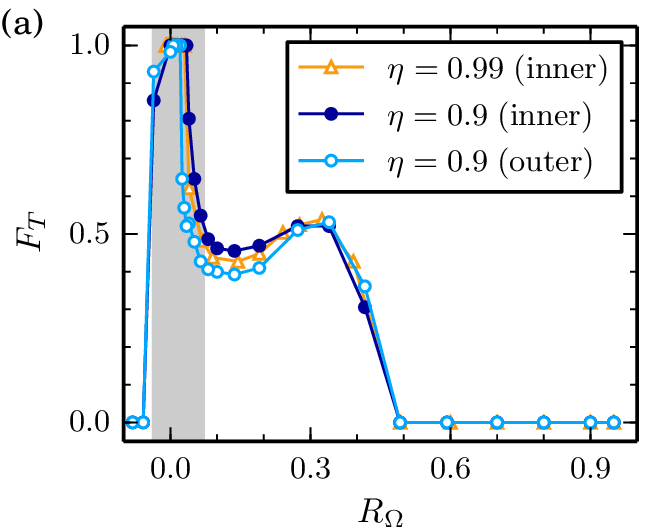}%
	            \includegraphics{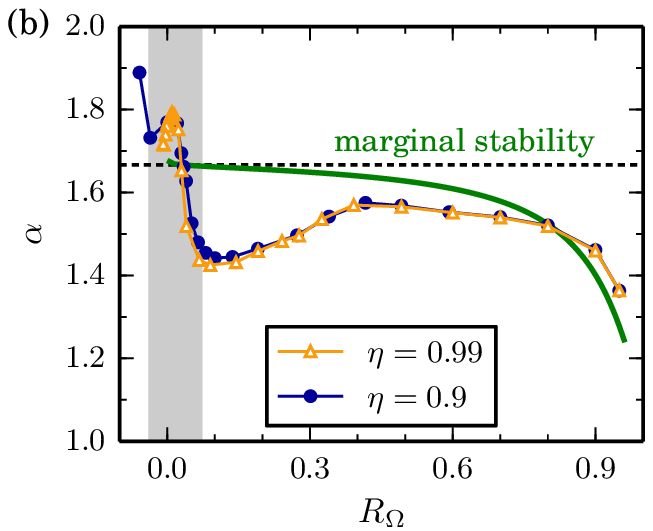}}
  \caption{(Colour online) Indicators for the transition to turbulent BLs in the DNS as a function of the rotation number $R_\Omega$: 
  (a)~Axial fraction $F_T$ covered by strongly turbulent regions with $\mathcal{L}_\mathrm{rms}>0.06$ (cf.~figure~\ref{fig:Lrms}) for $\Rey_S=2\e{4}$. 
  Since $F_T$ of the inner and outer BL coincide for $\eta=0.99$, only the inner value is shown. 
  (b)~The torque scaling exponent $\alpha$, defined by $G\propto \Rey_S^\alpha$, was calculated for each $R_\Omega$ individually 
  by a linear fit on a double-logarithmic scale using DNS results for $\Rey_S=5\e{3}$, $10^{4}$ and $2\e{4}$. 
  The corresponding scaling exponent for $\eta=0.99$ predicted by the marginal stability model (solid line) is based on model torques in the range $5\e{3}\leq \Rey_S \leq 2\e{4}$ 
  and approaches the asymptotic limit $\alpha=5/3$ (dotted line) for $R_\Omega \rightarrow 0$. 
  This marginal stability limit is exceeded in the DNS for $R_\Omega<0.03$.
  The BLs become turbulent for $R_\Omega<0.07$ in the region shaded in grey.}
  \label{fig:FT+scal}
\end{figure}
To analyse this transition process quantitatively, 
we calculate the axial fraction $F_T$ of each BL that is covered by strong turbulence with $\mathcal{L}_\mathrm{rms}>0.06$. 
Since the turbulent fractions of the outer and inner BL coincide for $\eta=0.99$, 
figure~\ref{fig:FT+scal}(a) only shows the latter for this radius ratio and both for $\eta=0.9$. 
For $R_\Omega\geq0.5$ no strongly turbulent BL region occurs in accordance with the small BL Reynolds number in this rotation-number range, cf.~figure~\ref{fig:ReBL}.
Then, in the range $0.07<R_\Omega<0.4$ the turbulent fraction increases, 
and approximately half of the BL becomes turbulent. 
This \textit{transitional regime} is also characterised by strong Taylor vortices \citep{Brauckmann2016a}, 
which interact with the BL dynamics \citep{Ostilla-Monico2014}. 
Finally, for $R_\Omega<0.07$ the turbulent fraction sharply increases to one, 
indicating the transition to fully turbulent BLs as assumed in the model. 
For $\eta=0.9$, the turbulent fraction drops again for negative rotation numbers, 
consistent with the fact that in these flow cases the outer cylinder strongly counter-rotates 
and thereby re-stabilises the flow.
Interestingly, the critical rotation number for the transition to turbulence depends on the wall curvature, 
whereas the general variation of $F_T$ with $R_\Omega$ does not differ between both $\eta$ values: 
For $\eta=0.9$, the inner and outer BL become turbulent at a larger and smaller $R_\Omega$ value, respectively, than the BLs for $\eta=0.99$. 
This difference represents another curvature effect 
and is consistent with the smaller inner (larger outer) transition Reynolds number $\Rey_{T,i}$ ($\Rey_{T,o}$), 
introduced to describe the DNS results for $\eta=0.9$ in~figure~\ref{fig:ReBL}(b).

As a second indicator for the BL transition, 
we analyse the local power-law scaling of the torque with $\Rey_S$, i.e. the exponent $\alpha$ from $G\propto \Rey_S^\alpha$, 
which was previously found to differ between flows with laminar and turbulent BLs 
\citep{Lathrop1992a,Lathrop1992,Lewis1999,Ostilla-Monico2014}. 
Based on the key assumption of laminar BLs that are marginally stable to the formation of Taylor vortices, 
a previous marginal stability calculation predicts the scaling exponent $\alpha=5/3$ in the limit of large $\Rey_S$ \citep{King1984,Marcus1984b}. 
The same exponent was calculated by \citet{Barcilon1984} by assuming BLs that are marginally stable to G\"{o}rtler vortices. 
In this context, a torque scaling exponent $\alpha>5/3$ has been linked to a flow with turbulent BLs \citep{Ostilla-Monico2014}. 

In~figure~\ref{fig:FT+scal}(b), we analyse the variation of the torque scaling exponent with $R_\Omega$ 
and compare the exponent from the DNS to the one predicted by the marginal stability model without BL transition.
We first note that there is hardly any difference between the cases of $\eta=0.9$ and $\eta=0.99$.
The exponent predicted by the model lies below the asymptotic value~$\alpha=5/3$ 
because it is calculated for finite $\Rey_S$ ranging between $5\e{3}$ and $2\e{4}$.
For~$R_\Omega\gtrsim 0.4$, the model qualitatively reproduces the variation of the exponent with $R_\Omega$ from the DNS, which also
suggests that the BLs are laminar in this regime. 
In the range $0.07<R_\Omega<0.4$ corresponding to the regime 
where laminar and turbulent regions in the BL coexist (cf.~figure~\ref{fig:FT+scal}a), 
the exponent significantly falls below the marginal stability prediction, 
as observed by \citet{Ostilla-Monico2014} in their \textit{transitional regime}.
Finally, for $R_\Omega<0.07$ the scaling exponent sharply rises demonstrating increasingly turbulent BLs. 
For $R_\Omega<0.03$, the exponent exceeds the marginal stability prediction $\alpha=5/3$ 
and accordingly indicates the torque scaling of a flow with completely turbulent BLs. 

Both the turbulent fraction $F_T$ and the scaling exponent $\alpha$ support the assumption that
the BLs become turbulent for $R_\Omega<0.07$.
This transition takes place in a small rotation-number range ($0.03<R_\Omega<0.07$), 
and it rationalizes the emergence of the narrow torque maximum at $R_\Omega=0.02$ with increasing $\Rey_S$.

\section{Summary and discussion}
The modelling of mean profiles from a turbulent flow using marginal stability arguments 
was previously successfully applied to thermal convection \citep{Malkus1954b} 
and to TC flow with stationary outer cylinder \citep{King1984,Marcus1984b,Barcilon1984}. 
While we here adopt the modelling arguments of \citet{King1984} and \citet{Marcus1984b}, 
some modifications were needed to generalise the marginal stability model to the case of independently rotating cylinders: 
As a first difference, the present model does not  
assume a constant angular momentum in the central region, 
and instead incorporates the small positive angular momentum gradient 
that was observed in simulations and experiments. While this brings 
the predictions closer to the DNS results for large $R_\Omega$, the effect of
the slope is not very big overall.
The second modification consists in the introduction of the increased effective gap widths 
$d_i=\widetilde{a}\delta_i$ and $d_o=\widetilde{a}\delta_o$, with $\widetilde{a}=1.5$, 
for the TC Reynolds numbers of the BLs. 
The constant $\widetilde{a}$ accounts for the enlarged space due to a free-surface-like 
boundary condition at the BL edge. 
Its value was kept fixed for all $\eta$ and $R_\Omega$. 
The previous model without the factor $\widetilde{a}$ underestimated the measured torques, 
as the comparison by \citet{Lathrop1992} showed.
Finally, as the marginal stability condition for both BLs, 
we here utilise an analytic formula that determines the TC stability boundary in the entire parameter space \citep{Esser1996}. 
In particular, this stability formula also applies to the wide-gap case
and to the situation of counter-rotating cylinders,
in contrast to approximate stability conditions used by \citet{King1984} and \citet{Chandrasekhar1961}.
However, for $\eta\rightarrow1$ and co-rotating cylinders, 
these approximations coincide with the stability boundary by \citet{Esser1996}
and therefore give the same results in this parameter range.

The simplifications of the model helped us to interpret the rotation dependence of the torque at $\Rey_S=2\e{4}$: 
While the broad maximum can be explained by marginal stability of both the central region and the BLs, 
the narrow torque maximum is related to turbulent BLs that enhance the angular momentum transport. 
Our simulations revealed that the assumption of laminar BLs is best justified for $R_\Omega\geq0.5$ 
and that a \textit{transitional regime}, where laminar and turbulent regions in the BL coexist, occurs for \mbox{$0.07<R_\Omega<0.4$}. 
The improved model that incorporates a shear transition suggests that 
the BLs become completely turbulent for $R_\Omega$ below  $0.07$. 
Remarkably, this demonstrates that the transition to turbulent BLs does not only depend on the shear strength ($\Rey_S$) 
as previously observed \citep{Lathrop1992,Lathrop1992a,Lewis1999,Ostilla-Monico2014}, 
but also on the system rotation, 
as also evidenced by the $R_\Omega$-dependence of the BL Reynolds numbers $\Rey_\mathrm{BL}^i$ and  $\Rey_\mathrm{BL}^o$. 
Moreover, the results for $\eta=0.9$ reveal that the transition to turbulent BLs depends on the wall curvature as well: 
At the convex inner cylinder, the transition occurs earlier 
(i.e. at a smaller critical value $\Rey_T$ and a larger~$R_\Omega$) than at the concave outer cylinder. 
We expect this curvature effect to become more pronounced for smaller radius ratios.

Previously, \citet{Lathrop1992} observed that the torque scaling exponent $\alpha=5/3$, 
predicted by marginal stability in the limit of large $\Rey_S$ \citep{King1984,Marcus1984b}, 
is incompatible with their torque measurements which show 
a continuous variation of $\alpha$ with $\Rey_S$ even in the regime of laminar BLs. 
Our calculations revealed that the exponent~$\alpha$ predicted by marginal stability also 
varies with $R_\Omega$ and $\Rey_S$ (not shown here) even at $\Rey_S$ as high as $10^4$.
As a consequence of this transitional behaviour, the predicted~$\alpha$ lies significantly below the asymptotic limit $\alpha=5/3$, 
in agreement with the torque computations in the regime of laminar BLs for $R_\Omega>0.5$.
The observed torque scaling exponent falls below the marginal stability prediction 
in the regime where laminar and turbulent regions in the BL coexist. 
It remains unclear how a mixture of laminar and turbulent BL regions can create 
a slower than laminar effective torque scaling. 
However, the observed $\alpha$ clearly exceeds the marginal stability limit $5/3$ 
in the regime where the BLs are completely covered with turbulence.

We here investigated the torque and BL behaviour in low-curvature TC flow for $\eta\geq 0.9$. 
We expect that the application of the marginal stability model to TC flow with $\eta<0.9$ underlies some limitations.
Already the larger discrepancies between model and DNS for $\eta=0.9$ suggest 
that curvature effects become relevant for $\eta<0.9$. 
More importantly, the marginal stability model does not account for the intermittent bursting behaviour 
observed in the outer flow region for strongly counter-rotating cylinders \citep{VanGils2012,Brauckmann2013a}: 
Since the flow switches over time between quiescent and highly turbulent phases, 
the assumption of one marginally stable state is inadequate here. 
The intermittent behaviour gains in importance with decreasing~$\eta$, 
because the bursting becomes stronger \citep{Brauckmann2016a} and occurs in a wider rotation-number range 
since the regime of counter-rotating cylinders corresponds to $-(1-\eta)/\eta < R_\Omega < 1-\eta$.

Finally, the model predicts that the BL Reynolds numbers increase with $\Rey_S$,
and therefore the $\Rey_\mathrm{BL}(R_\Omega)$ curves in~figure~\ref{fig:ReBL} shift upwards for $\Rey_S>2\e{4}$. 
Then, the predicted BL Reynolds numbers exceed the transition threshold $\Rey_T$ already at a larger critical $R_\Omega$ value, 
and thus the BLs become turbulent in a wider rotation-number range.
Therefore, we expect that the broad torque maximum caused by marginal stability will disappear at higher $\Rey_S$. 
Moreover, in the simulations, the narrow torque maximum at~$R_\Omega=0.02$ grows faster with $\Rey_S$ (cf.~figure~\ref{fig:FT+scal}b), 
which also suggests that it will eventually outperform the broad maximum. 
Torque measurements for $\eta=0.909$ and $\Rey_S$ of a few $10^5$ \citep{Ostilla2014} support this expectation 
and show only a single maximum at $R_\Omega=0.04$ close to the value $R_\Omega=0.02$ found here.\\

%
%
We thank D. P. Lathrop and D. Lohse for fruitful discussions, 
M. Avila for providing the code used for the TC simulations, 
and the LOEWE-CSC at Frankfurt for access to its computational resources.
This work was supported in part by the DFG via Forschergruppe 1182 Wandnahe Prozesse.

\bibliographystyle{jfm}

\end{document}